\documentclass{article}
\usepackage{color}

\newcommand{\omits}[1]{}
\newcommand{\be}{\begin{equation}}
\newcommand{\ee}{\end{equation}}
\newcommand{\beqa}{\begin{eqnarray}}
\newcommand{\eeqa}{\end{eqnarray}}
\newcommand{\beqar}{\begin{eqnarray*}}
\newcommand{\eeqar}{\end{eqnarray*}}

\newcommand\sech{{\rm sech}}
\definecolor{dyellow}{rgb}{1.,0.8,.0}
\definecolor{myblue}{rgb}{.1,.1,.7}
\definecolor{dcyan}{rgb}{.0,.6,.6}
\definecolor{dmagenta}{rgb}{0.6,0.0,0.6}
\definecolor{brown}{rgb}{0.6,0.2,0.}
\definecolor{darkblue}{rgb}{.0,.0,0.5}
\definecolor{darkred}{rgb}{0.75,0.0,0.0}
\definecolor{orange}{rgb}{1.,.6,.0}
\definecolor{dorange}{rgb}{0.8,.4,.0}
\definecolor{lightgray}{rgb}{0.7,0.7,0.7}
\definecolor{darkgreen}{rgb}{0.0,0.6,0.0}
\definecolor{purple}{rgb}{.4,.0,.4}

\def\PRD{{Phys. Rev.}~{\bf D}}
\def\PRL{{Phys. Rev. Lett. }}

\def\CQG{{Class. Quant. Grav. }}
\def\CMP{{Commun. Math. Phys. }}

\def\PRP{{Phys. Rep.}}

\def\MPLA{{Mod. Phys. Lett.}~{\bf A}}
\def\IJMPD{{Int. J. Mod. Phys.}~{\bf D}}

\begin{document}

\centerline{\Large \bf Thermodynamic Properties of Spherically-Symmetric,} \vskip 0.15cm
\centerline{\Large \bf Uniformly-Accelerated Reference Frames}


\bigskip

\centerline{\bf Chao-Guang Huang\footnote{huangcg@ihep.ac.cn}
and Jia-Rui Sun\footnote{sun@ihep.ac.cn},} \vskip .3cm
\centerline{\it Institute of High Energy Physics, Chinese Academy
of Sciences,} \centerline{\it P.O. Box 918-4, Beijing 100049,
China}

\begin{abstract}
We aim to study the thermodynamic properties of the spherically
symmetric reference frames with uniform acceleration, including
the spherically symmetric generalization of Rindler reference
frame and the new kind of uniformly accelerated reference frame.
We find that, unlike the general studies about the horizon
thermodynamics, one cannot obtain the laws of thermodynamics for
their horizons in the usual approaches, despite that one can
formally define an area entropy (Bekenstein-Hawking entropy). In
fact, the common horizon for a set of uniformly accelerated
observers is not always exist, even though the Hawking-Unruh
temperature is still well-defined. This result indicates that the
Hawking-Unruh temperature is only a kinematic effect, to gain the
laws of thermodynamics for the horizon, one needs the help of
dynamics. Our result is in accordance with those from the various
studies about the acoustic black holes.

\bigskip
\noindent PACS number 04.70.Dy
\newline
\end{abstract}

\bigskip


\section{Introduction}

People used to studying physics in the inertial reference frame.
However, when the accelerated reference frame is involved, the
physics becomes rather different. A well-known example is the
vacuum state in a flat spacetime. The vacuum state of a quantum
field in a flat spacetime is independent of the choice of inertial
frames due to the Pioncar\'e symmetry. Nevertheless, an
accelerated detector will register particles in the vacuum of
Minkowski spacetime, i.e., detect a background temperature of the
flat spacetime \cite{unruh1}, which indicates that the concept of
quantum state is dependent of the choice of reference frames.
Since the Rindler reference frame 
\cite{rindler1} has a temperature, it is naturally to study its
thermodynamic properties. Some works have been done on this topic
\cite{pd1,hc} and their results incline to the existence of
horizon entropy and the laws of thermodynamics for the Rindler
horizon. The aim of the present paper is to study whether there
exists thermodynamics for the horizon of the reference frame with
uniform acceleration. Now that the definition of entropy for the
horizon needs the horizon area be finite, we focus on discussing
the thermodynamic properties of the spherically symmetric
generalization of both Rindler reference frame and the new kind of
uniformly accelerated reference frame recently proposed in
\cite{CG}. Both cases describe a set of uniformly accelerated
observers in which all the observers initially locate at the same
sphere with some fixed radius, and move (outward or inward) along
the radius with the same 4-velocity and 4-acceleration. When the
set of observers moving outward, each of them has his own event
horizon. Each 2-dimensional horizon is a plane which divides the
space into two parts: one has causal connection to the observer,
another does not have, corresponding to the causal or noncausal
part, respectively. The intersection of all the noncausal parts is
a 3-dimensional ball in space. Its boundary, a sphere, just acts
as the 2-dimensional common event horizon for these observers,
i.e., all of them locate outside the sphere cannot `see' the
inside of the sphere. Viewed in this way, we define it as the
2-dimensional event horizon for the set of uniformly accelerated
observers. In addition, one may formally define an area entropy
for it. When the set of observers moving inward, each of them also
has his own event horizon, and each 2-dimensional horizon is again
a plane dividing the space into two parts. But the intersection of
those noncausal parts is now empty! So there is no common event
horizon for this set of uniformly accelerated observers. In other
words, the event horizon for this set of observers does not exist.
If one persisted in introducing an area entropy, one immediately
find that the area entropy becomes ill-defined. Meanwhile, one
should note that the Hawking-Unruh temperature is always
well-defined in all the situations. That is to say, one cannot
assign an area entropy to the event horizon for the set of
uniformly accelerated observers because the horizon is not always
exist in these reference frames. The studies about the spherically
symmetric generalization of Rindler reference frame and the new
kind of uniformly accelerated reference frame both indicate that
the Hawking-Unruh temperature is only a kinematic effect, to gain
the laws of thermodynamics for the event horizon, one needs the
help of dynamics. This point is in accord with the previous
studies about the acoustic black holes
\cite{unruh,visser1,visser2}, and gives examples that are
different to the general horizon thermodynamics (e.g., Rindler
spacetime, de Sitter spacetime and Friedmann-Robertson-Walker
(FRW) universe) which have been widely studied by many authors
\cite{pd1,hc,GH,Strominger,hlw,ghz,ck,pd}.

The arrangement of this paper is in the following way: we begin in
the next section to give a brief review to the Rindler reference
frame, M$\o$ller reference frame and a new kind of reference frame
with uniform acceleration. In section {\bf 3}, we pay our special
attention to their spherically symmetric generalizations and study
their thermodynamic properties. Finally, we will make the
conclusions and discussion.

Throughout the paper, we consider the 4-dimensional spacetime and use the
natural units $\hbar=c=G=k_B=1$.

\section{Unifomly accelerated reference frames }
\subsection{Rindler and M$\o$ller reference frame}

The Rindler reference frame 
is often expressed in
Cartesian coordinates \cite{rindler1,Rindler},
\be
 ds^2=-x^2dt^2+dx^2+dy^2+dz^2.
\ee
For later convenience, we make a little change of the above metric
\be \label{rind}
 ds^2=-g^2x^2dt^2+dx^2+dy^2+dz^2  , \ee
where $g$ is a constant. A static observer with a 4-velocity
$U^\mu=\{1/(gx),0,0,0\}$ possesses a 4-acceleration
$a^{\mu}=U^\nu\nabla_{\nu}U^\mu=\{0,1/x,0,0\}$. Its magnitude is
$a\equiv(a^{\mu} a_{\mu})^{1/2}=1/x$. It is clear to see that at
each fixed space point $\{x,y,z\}$, the acceleration is a
constant, but the acceleration varies point by point, namely,
different observer possesses different acceleration.

By making the coordinates transformations of the Rindler reference
frame
\be \label{R2M}
t\rightarrow \tilde{t}\quad {\rm and} \quad x\rightarrow
(1+g\tilde{x})/g.
\ee
Eq.(\ref{rind}) changes to the M$\o$ller reference frame 
\cite{MTW}
\be \label{moller}
 ds^2=-(1+g\tilde{x})^2d{\tilde{t}}^2+d{\tilde{x}}^2+dy^2+dz^2.
\ee
\omits{s}Similar calculation shows that the magnitude of the
4-acceleration for a static observer with $U^\mu=\{1/(1+g\tilde
x),0,0,0\}$ at fixed $\tilde{x},y,z$ is
\be
 a=\frac{g}{1+g\tilde{x}}
\ee
which is also a constant. However, different static observer
possesses different acceleration in M$\o$ller reference frame,
too.

The Rindler horizon has a temperature $T_H$ relating to its
`surface gravity'
\be \label{kappa}
\kappa=\lim_{x\to x_H}Va=g
\ee
by the known relation $T_H = \kappa /2\pi$ between surface gravity
and Hawking-Unruh temperature,
\be \label{tem}
T_H=\frac{g}{2\pi},
\ee
where $V=\sqrt{-g_{00}}$ is the redshift factor. The Hawking-Unruh
temperature corresponding to the M$\o$ller reference frame is the
same as Rindler's.

\subsection{New uniformly accelerated reference frame}

The new kind of uniformly accelerated reference frame was first
proposed in \cite{CG} in Cartesian coordinates
\be
\label{cartisian}%
ds^2=-dT^2+2\sinh(aT)dTdX+dX^2+dy^2+dz^2, \ee
where $a$ is a constant, for a `static' observer with 4-velocity
$U^\mu=\{1,0,0,0\}$, it is easy to see that his 4-acceleration is
$a^\mu=\{a \tanh(aT),a\, \sech(aT),0,0\}$, and the magnitude of
$a^\mu$ is just $(a^{\mu} a_{\mu})^{1/2}=a$ which is indeed a
constant acceleration. In addition, it has been shown in \cite{CG}
that the equation of motion reduces to the standard second law of
mechanics in the Newtonian approximation, whereas, the second law
of mechanics in Rindler reference frame does not have this
property.

By a Wick rotation of the time coordinate
\be
 T\rightarrow\tau=iT
\ee
in Eq.(\ref{cartisian}), we get the imaginary time period $\beta=
2\pi /a$, which leads to the Hawking-Unruh temperature of the
horizon
\beqa
 T_H=\frac 1 \beta=\frac{a}{2\pi}.
\eeqa
If one uses the relation between temperature and `surface
gravity', he or she will get the same Hawking-Unruh temperature.

\section{Thermodynamic Properties of Spherically Symmetric Reference Frames}

As we have mentioned in Introduction, the planar Rindler reference
frame 
and the new kind of uniformly accelerated
reference frame is not very convenient for studying their
thermodynamic properties, e.g., defining the entropy for the
horizon requires the horizon to have a finite surface. So we will
generalize these uniformly accelerated reference frames in section
{\bf 2} into spherically symmetric cases.

\subsection{Spherically symmetric generalization of Rindler reference frame}

Firstly, we study the generalized Rindler reference frame. Recall
that the 4-dimensional Minkowski spacetime in spherical
coordinates is
\be \label{msp}
 ds^2=-dt^2+dr^2+r^2(d\theta^2+\sin^2\theta
d\phi^2).
\ee
After using the following coordinates transformations
\be \label{rspc} \rho-\rho_0=\pm[(r-r_0)^2-t^2]^{1/2}\quad{\rm
and}\quad \eta=\frac{1}{a}\tanh^{-1}[t/(r-r_0)], \ee
the line-element (\ref{msp}) becomes
\be \label{rsp}
 ds^2=-a^2(\rho-\rho_0)^2d\eta^2+d\rho^2+r^2(\eta,\rho)
(d\theta^2+\sin^2\theta
d\phi^2), \ee
where $r(\eta,\rho)$ means $r$ is the function of $\eta$ and
$\rho$. A simple calculation shows that Eq.(\ref{rsp}) is just the
generalized Rindler reference frame in spherically symmetric
coordinates, since the 4-acceleration is
$a^{\mu}=\{0,(\rho-\rho_0)^{-1},0,0\}$ for a static observer with
4-velocity $U^\mu=\{ a^{-1}(\rho-\rho_0)^{-1},0,0,0\}$, and the
corresponding Hawking-Unruh temperature is $T_H=a/(2\pi)$. This
reference frame describes a set of observers locate at the same
sphere with some fixed radius $r$, and all of them move (outward
or inward) along the radius with the same $U^{\mu}$ and $a^{\mu}$.
It can also be seen from the calculation that the `+' in
Eq.(\ref{rspc}) corresponds to observers moving outward along the
$r$ axis with uniformly acceleration, while `-' corresponds to the
inward moving case. Note that each observer has his own event
horizon, and like the planar Rindler case, each 2-dimensional
horizon is a 
plane tangent to the sphere
$\rho=\rho_0$. To find the common event horizon for this set of
uniformly accelerated observers, we study the following two
situations.

When this set of uniformly accelerated observers moving outward
along $r$ axis, one has $r=r_0+t$. For each observer, there is a
2-dimensional event horizon which divides the space into a causal
part and a noncausal part. The intersection of all these noncausal
parts form a 3-dimensional ball, light signals emitted in the ball
cannot affect anyone of the observers since these observers locate
at some $r>r_0$ (i.e., outside the sphere). That is to say, the
boundary of the 3-dimensional ball just acts as the common event
horizon for these observers, namely, they have the same event
horizon. Thus we define the 2-dimensional sphere as the event
horizon for this set of uniformly accelerated observers. If the
Bekenstein-Hawking entropy
\be \label{eor}
 S\omits{=\frac{1}{4}4\pi r^2|_{\rho=\rho_0}}=\pi(r_0+
t)^2 \ee
is assigned for the horizon, it seems 
not to violate the second law of thermodynamics as the
time $t$ evolves.

When these observers moving inward to the origin $r=0$ along $r$
axis with uniform acceleration $a^\mu$, $r=r_0-t$. One should pay
attention that in this situation, although for each observer there
is an event horizon, for all the observers, there will be no
common event horizon at all! This is because the spatial section
of the event horizon for each observer is a 2-dimensional plane
tangent to the sphere at $\rho=\rho_0$ (i.e., $r=r_0-t$), which
separates the space into two parts. In the causal part, a light
signal emitted in this region can affect the observer eventually.
While in the noncausal part, no signal can affect the observer.
The intersection of these noncausal parts is empty.  However, the
`envelop' of these 2-dimensional plane at the same sphere is also
a 
sphere, if its radius $r=r_0-t$ is positive. Notice that all of
the uniformly accelerated observers locate at some $r<r_0-t$ since
$\rho<\rho_0$ in this case, that is to say, these observers are
within the sphere, which indicates that light signals emitted from
any place can always affect some observers. Namely, the
2-dimensional sphere cannot be regarded as the common event
horizon for this set of uniformly accelerated observers.
Consequently, there is no such an area entropy for the
horizon.\omits{This case is illustrated in Fig.1} If one persisted
in assigning an area entropy to the 2-dimensional sphere by
analogy with the black hole thermodynamics, i.e., the entropy $S$
was a quarter of the area $A$ of the sphere, where $A=4\pi
r^2|_{\rho=\rho_0}=4\pi(r_0- t)^2$ ($r_0$ is a constant), the
entropy
\be
 S=\omits{\frac{1}{4}4\pi r^2|_{\rho=\rho_0}=}\pi(r_0- t)^2
 \ee
would decrease for the isolated system as the time $t$ evolves
when $r_0>0$. Obviously, it will violate the second law of
thermodynamics. Thus, the entropy of horizon for inward, uniformly
accelerated observers is ill-defined.

\subsection{Spherically symmetric generalization of the new uniformly accelerated reference
frame }

Now, we turn to the generalized new uniformly accelerated
reference frame. By making the following coordinates
transformations
\be \label{ctn}
 t=\frac{1}{a}\sinh(aT)\quad{\rm and}\quad
r=R+\frac{1}{a}(\cosh(aT)-1)
\ee
in Eq.(\ref{msp}), we get a new spherically symmetric coordinates
in the Minkowski spacetime
\be
\label{spherical}
ds^2=-dT^2+2\sinh(aT)dTdR+dR^2+[R+\frac{1}{a}
(\cosh(aT)-1)]^2(d\theta^2+\sin^2\theta d\phi^2).
\ee
It can be seen that the first three terms in (\ref{cartisian}) and
(\ref{spherical}) are quite similar. This is because the same form
of coordinates transformations has been used as in getting
(\ref{cartisian}) from the flat Minkowski spacetime. A short
calculation shows that the 4-acceleration is $a^\mu=\{a\tanh(aT),a \,
\sech(aT),0,0\}$ for a `static' observer with 4-velocity
$U^\mu=\{1,0,0,0\}$, and the magnitude of the acceleration
$(a^{\mu} a_{\mu})^{1/2}=a$ is a constant. Therefore, this
reference frame just describes a set of uniformly accelerated
observers in which all of them initially locate at the same sphere
with some fixed radius, and move (outward or inward) along the
radius with the same $U^\mu$ and $a^{\mu}$.

Since this uniformly accelerated reference frame is described by a
metric which is neither static nor stationary, the usual approach
to determine the location of its  horizon like for the Rindler
reference frame and Schwarzschild black hole is not applicable.
Instead, we search it following the approach in cosmology. Note
that the 3-dimensional event horizon is generated by the null
curves. That is, we have
\be
ds^2=0
\ee
for the generators of the horizon. When the set of observers on a
sphere with radius $R$ and moving towards the radial direction,
respectively. Eq.(\ref{spherical}) becomes
\be
ds^2=-dT^2+2\sinh(aT)dTdR+dR^2=0,
\ee
which can be rewritten as
\be
\frac{dR^2}{dT^2}+2\sinh(aT) \frac{dR}{dT}-1=0,
\ee
this differential equation has two solutions
\be
\frac{dR}{dT}= \exp[-aT] \quad{\rm or}\quad \frac{dR}{dT}=-
\exp[aT].
\ee
The first solution indicates the future outgoing mode, while the
second one indicates the future incoming mode. Consider these
observers locate at large enough $R$, one has 
\be
 R-R_H=\int_{R_H}^R dR=\int_T^\infty
\exp[-aT]dT=\frac{1}{a}\exp[-aT]. \ee
Namely, the 2-dimensional future event horizon is at
\be \label{h}%
 R_H=R-\frac{1}{a}\exp[-aT]
\ee
in the uniformly accelerated reference frame. In particular, at
$T=0$,
\be \label{h0}%
R_H=R-\frac{1}{a}.
\ee
While $T\to \infty$, the event horizon will reach its maximum

\be
\lim_{T\to \infty}R_H\rightarrow R_m=R,
\ee
i.e., approach to the position of observer. Obviously, the sets of
observers at different radius have different horizons!

The horizon area $A$ is
\beqa \label{area}
A &=& \int \sqrt{\sigma}d\theta d\phi =
4\pi [R_H + \frac{1}{a} (\cosh(aT)-1)]^2  \nonumber \\
&=& 4\pi [R+\frac{1}{a}(\sinh(aT)-1)]^2,
\eeqa
where $\sigma$ is the determinant of the induced metric on the
2-dimensional sphere.

The Hawking-Unruh temperature of the horizon is easily obtained
from the Wick rotation: $T_H = a/(2\pi)$. One may also formally
define a Bekenstein-Hawking entropy $S$ for the event horizon by
analogy with the black hole entropy, i.e.,
\be 
 S=\frac{A}{4} \omits{.
\ee
Substituting Eq.(\ref{area}) into
Eq.(\ref{entropy}), we get the explicit form of the entropy
\be
 S}=\pi[R+\frac{1}{a}(\sinh(aT)-1)]^2.\label{eou}
\ee

Nevertheless, according to the experience we learned from the
spherically symmetric generalization of Rindler reference frame,
one should take care whether it make sense to assign such an
entropy to the event horizon of this new reference frame. Recall
that $R$ which comes from the coordinate transformations in
Eq.(\ref{ctn}) is a constant for some given `static' observers in
the uniformly accelerated reference frame. When
$R<a^{-1}\exp[-aT]$ or $R<a^{-1}$, Eq.(\ref{h}) or Eq.(\ref{h0})
gives the negative value. In both situations, light signals
emitted from any place can always affect some observers on the
sphere with radius $R$, which means that in these situations,
there are no common event horizon for the set of observers.
Consequently, no area entropy could be introduced. However, the
Hawking-Unruh temperature is always exist for accleratad observers
based on Unruh's discussion on the thermal effect of an
accelerated detector \cite{unruh1}.

The absence of event horizon but meanwhile the existence of
Hawking-Unruh temperature is a new property of the spherically
symmetric accelerated reference frames including the
generalization of Rindler reference frame and the generalization
of this new reference frame with `dynamic' metric. This gives
different examples to the general horizon thermodynamics which
have been widely studied by many authors
\cite{hc,GH,Strominger,hlw,ck,pd\omits{,SUN}}.

\subsection{Hawking-Unruh temperature without area entropy}

What do they tell us? Following the above discussions, both of the
two examples: generalized Rindler reference frame and generalized
new uniformly accelerated reference frame, seem to give us a hint
that the thermodynamic properties of their horizons may be
different from those of the general horizon thermodynamics, like
the black holes, de Sitter spacetime and FRW universe, in which
the Einstein equation is used in getting the laws of
thermodynamics. Pay attention that, in obtaining the
Hawking{-Unruh} temperature for the two examples, one indeed does
not apply any dynamics at all. Moreover, the Hawking-Unruh
temperature is always well-defined even though the common event
horizon of the set of uniformly accelerated observers does not
exist. So these two examples, both indicate that the appearance of
the Hawking-Unruh temperature (or Hawking radiation) is only a
kinematic effect but not a dynamic one. This conclusion is in
accord with the studies about the acoustic black holes
\cite{unruh,visser1,visser2}, (for a review, see \cite{visser1}).
Where in these cases, on one hand, by making an analogy between
the sonic field in fluid field and the quantum field in
gravitational field, a kind of acoustic metrics which are called
the acoustic black holes are obtained, they are of the same form
of the Schwarzschild metrics, and then the Hawking radiation is
obtained naturally. On the other hand, the fluid flow system has
nothing to do with the real black hole system, hence the
introduction of entropy to their acoustic horizon is meaningless.
In turn, one can learn from these examples that:

\smallskip

\noindent $a$. the Hawking radiation is only a kinematic effect, regardless
of the dynamics, e.g., the Einstein equation (which, in fact,
can have been seen from Hawking's original derivation
\cite{hawk1}\cite{hawk2}); and

\smallskip
\noindent $b$. it is only when the dynamics is applied, e.g., the Einstein
equation, that one can assign the entropy to the horizon, and
hence gain the laws of thermodynamics for the horizon (which can be
seen from \cite{BCH} and has been proved indirectly in \cite{jacobson}
to some degree).

In addition, studies about acoustic black holes indicate that the
Hawking radiation (or Hawking-Unruh temperature) will exist in any
Lorentzian geometry with an event horizon \cite{visser1,visser2},
but according to the discussion in previous subsections, our
examples go further: Hawking-Unruh temperature will exist even
when there is no common event horizon for the set of observers
with uniform acceleration (though each observer, of course, has his own
event horizon).

So far, we have answered the question which we have previously
proposed, that is, the existence of Hawking-Unruh temperature is
only a kinematic effect, to obtain the laws of thermodynamics for
the horizon, one needs the help of dynamics. Since the spherically
symmetric generalizations of the Rindler reference frame and the
new uniformly accelerated reference frame indeed do not involve
any dynamics, so assigning the entropy and laws of thermodynamics
for their event horizons (if the common event horizon exists) make
no sense.

\section{Conclusions and Discussion}
Through the study of thermodynamic properties of the
generalizations of Rindler reference frame and a new kind of
uniformly accelerated reference frame in spherically symmetric
coordinates case, we finally find that there is no laws of
thermodynamics for them by applying the usual approaches. Our
result, not only supports the studies of acoustic black holes, but
also broaden their conclusions to the situation in which the
common event horizon is not always exist for the set of uniformly
accelerated observers. All of this, seems to indicate one thing
that the Hawking-Unruh temperature, which is only a kinematic
effect, has no intrinsic connection with the horizon area entropy
(Benkenstein-Hawking entropy), which is a holographic property
corresponding to the gravitation. Hence in order to get the
thermodynamic laws and further find the microscopic origin of the
Benkenstein-Hawking entropy for black holes, the gravitational
laws, i.e., the Einstein equation or other quantum theory of
gravity should be introduced.

In the end, review that when the initial condition is properly
selected to let the horizon make sense, we can see that e.g., the
entropy $S$ in Eq.(\ref{eou}) is never decreasing as the time $T$
evolves when $R_H$ in Eq.(\ref{h0}) is positive, which implies
that something like the generalized second law of thermodynamics
is hold for this system \cite{bekenstein1,bekenstein2}. Be that as
it may, we prefer to take this as merely an analogy but not of the
real physical meaning.

\vspace{0.2in}

\paragraph{Acknowledgments:}

We are grateful to Prof. H.-Y. Guo for stimulating discussions.
JRS would like to thank Prof. R.-G. Cai and X. Li for useful
discussions and comments, and thank T. Qiu for making comment on
the earlier draft of this paper. This work is supported by NSFC
under Grant Nos. 90403023 and 10375087.



\begin{thebibliography}{10}

\bibitem{unruh1}
W. G. Unruh,
\newblock ``Notes on black-hole evaporation,''
\newblock \PRD {\bf 14}, 870 (1976).



\bibitem{rindler1}
W.~Rindler,
\newblock ``Kruskal space and the uniformly accelerated frame,''
\newblock Am. J. Phys. {\bf34}, 
1174 (1966).



\bibitem{pd1}
T.~Padmanabhan,
\newblock ``Classical and quantum thermodynamics of
horizons in spherically symmetric spacetimes''
\newblock gr-qc/0204019.
\newblock ``Thermodynamics of horizons: A comparison of
Schwarzschild, Rindler and de Sitter spacetimes,''
\newblock \MPLA 17, 923 (2002).



\bibitem{hc}
H. Culetu,
\newblock ``Is the Rindler horizon energy nonvanishing?''
\newblock hep-th/0607049.



\bibitem{CG}
C.-G. Huang and H.-Y. Guo,
\newblock ``A new kind of uniformly accelerated reference frame",
\newblock \IJMPD {\bf 15}, 1035 (2006), gr-qc/0604008.



\bibitem{unruh}
W. G. Unruh,
\newblock ``Experimental black-hole evaporation?''
\newblock \PRL {\bf 46}, 1351 (1981).



\bibitem{visser1}
M. Visser,
\newblock ``Acoustic black holes: Horizons, ergospheres, and Hawking
radiation''
\newblock \CQG {\bf 15}, 1767 (1998), gr-qc/9712010.



\bibitem{visser2}
M. Visser,
\newblock ``Hawking radiation without black hole entropy''
\newblock \PRL {\bf 80}, 3436 (1998), gr-qc/9712016.



\bibitem{GH}
G. W.~Gibbons, S. W. Hawking,
\newblock ``Cosmological event horizons, thermodynamics, and particle
creation''
\newblock \PRD {\bf 15}, 2738 (1977).




\bibitem{Strominger}
M. Spradlin, A. Strominger and A. Volovich,
\newblock `` Les Houches lectrues on de Sitter space''
\newblock hep-th/0110007.




\bibitem{hlw}
C.-G.~Huang, L. Liu and B. Wang,
\newblock ``Thermodynamics of de Sitter universes''
\newblock \PRD {\bf 65}, 083501 (2002).




\bibitem{ghz}
H.-Y. Guo, C.-G. Huang and B. Zhou,
\newblock `` Temperature at horizon in de Sitter spacetime''
\newblock Europhys. Lett.{\bf 72}, 1045 (2005), hep-th/0404010.




\bibitem{ck}
R.-G. Cai, S. P. Kim,
\newblock `` First law of thermodynamics and Friedmann equations of Friedmann-Robertson-Walker
universe''
\newblock JHEP 0502, 050 (2005), hep-th/0501055.




\bibitem{pd}
T. Padmanabhan,
\newblock ``Cosmological constant: The weight of the vacuum''
\newblock \PRP\ {\bf 380}, 235 (2003), hep-th/0212290;
\newblock ``Gravity and the thermodynamics of horizons''
\newblock \PRP\ {\bf 406}, 49 (2005), gr-qc/0311036, and related references
therein.




\bibitem{Rindler}
W. Rindler,
\newblock {\it Essential Relativity, --- special, general, and
cosmological}, 2nd Edition, (Springer-Verlag, New York, 1977)''.




\bibitem{MTW} C. M{\o}ller, {\it The theory of relativity},
(Oxford university Press, London, 1952);
C. W. Misner, K. S. Thorne and J. A. Wheeler,
\newblock {\it Gravitation},(Freeman,
San Francisco, 1973).''



\bibitem{hawk1}
S. W. Hawking,
\newblock ``Black hole explosions?''
\newblock Nature {\bf 248}, 30 (1974).



\bibitem{hawk2}
S. W. Hawking,
\newblock ``Particle creation by black holes''
\newblock \CMP {\bf 43}, 199 (1975).



\bibitem{BCH}
J. M. Bardeen, B. Carter, and S. W. Hawking,
\newblock ``The four laws of black hole mechanics''
\newblock \CMP {\bf 31}, 161 (1973).



\bibitem{jacobson} T. Jacobson,
\newblock ``Thermodynamics of space-time: The Einstein equation of
state''
\newblock \PRL {\bf 75}, 1260 (1995), gr-qc/9504004.



\bibitem{bekenstein1}
J. D.~Bekenstein,
\newblock ``Black  holes and entropy'',
\newblock \PRD {\bf 7}, 2333 (1973).



\bibitem{bekenstein2}
J. D. Bekenstein,
\newblock ``Generalized second law of thermodynamics in black-hole
physics''
\newblock \PRD {\bf 9}, 3292 (1974).





\end{thebibliography}
\end{document}